\documentclass[11pt]{amsart}

\usepackage{latexsym}
\usepackage{amsfonts}
\usepackage{amsthm}
\usepackage{graphicx}
\usepackage{amssymb}
\usepackage{amsmath}
\usepackage{amssymb}
\usepackage{color}
\author[]{Dima Grigoriev}
\address{CNRS, Math\'ematiques, Universit\'e de Lille, 59655, Villeneuve d'Ascq, France}
\email{dmitry.grigoryev@math.univ-lille1.fr}

\author[]{Vladimir Shpilrain}
\address{Department of Mathematics, The City  College  of New York, New York,
NY 10031} \email{shpil@groups.sci.ccny.cuny.edu}
\thanks{Research of the second author was partially supported by
the NSF grant DMS-0405105. }

\dedicatory{Dedicated to Ben Fine on his 60th birthday}

\begin{document}

\title[]{Authentication from matrix conjugation}

\begin{abstract}
We propose an authentication scheme where forgery (a.k.a.
impersonation) seems  infeasible  without finding the prover's
long-term private key. The latter would follow from solving the
conjugacy search problem in the platform (noncommutative) semigroup,
i.e., to recovering $X$ from $X^{-1}AX$ and $A$. The platform
semigroup that we suggest here is the semigroup of $n \times n$
matrices over truncated multivariable polynomials over a ring.

\end{abstract}

\maketitle

\section{Introduction}

For a general theory of public-key authentication (a.k.a.
identification) as well as early examples  of authentication
protocols, the reader is referred to \cite{Menezes}. In this paper,
we propose an authentication scheme where recovering the private key
from the public key would follow from   solving the conjugacy search
problem in the platform (noncommutative) semigroup, i.e., to
recovering $X$ from $X^{-1}AX$ and $A$. There were some previous
proposals based on this problem, see e.g. \cite{KCCL, SDG}, so it
would make sense to spell out what makes our proposal different:

\begin{enumerate}

\item Forgery (a.k.a.  impersonation) seems infeasible without finding the prover's long-term  private key.
In other proposals, there is usually a ``shortcut", i.e., a way for
the adversary to pass the final test by the verifier without
obtaining the prover's private key. In particular, in the proposal
of \cite{KCCL} modeled on the Diffie-Hellman  authentication scheme,
there is an alternative (formally weaker) problem that is sufficient
for the adversary to solve in order to impersonate the prover.
Namely,   it is sufficient for the adversary to obtain
$Y^{-1}X^{-1}AXY$ from $X^{-1}AX$, $Y^{-1}AY$,  and $A$.

\item  Our platform semigroup might be the first serious candidate for having generically
hard conjugacy search problem. It can
therefore be used with some other previously suggested cryptographic
protocols based on the conjugacy search problem, e.g. with the
protocols in \cite{AAG} or \cite{Grigoriev}; see also \cite{ourbook}
for more examples.

\item  One of the most important new features is that the verifier
{\sl selects his final test randomly from a large
series of tests}. This is what makes it difficult for the adversary
to  impersonate the prover without obtaining her private key: if the
adversary just ``studies for the test", as weak students do, he/she
at least should know what the test is.

\item Unlike the proposals in \cite{GS, SDG}, our authentication scheme does not use the
Feige-Fiat-Shamir idea \cite{Fiat} involving
repeating several times a three-pass challenge-response step (to avoid predicting,
by the adversary, the challenge with
non-negligible probability). In our scheme, we have just one challenge and one response.

\item  To prevent attacks by {\it malicious verifier}, there is an  intermediate ``commitment to challenge" step
for the verifier because otherwise, malicious verifier might present
the prover with a carefully selected challenge that may result in
leaking information about the prover's private key at the response
step. This is similar to the ``chosen-plaintext attack" on an
encryption protocol.


\end{enumerate}

Perhaps it is worth spelling out that in this paper, our main focus
is on how to protect  the prover's long-term private key from any
information leaks during authentication sessions. We put less
emphasis here on security of the prover's long-term private key
against attacks on her long-term public key.

\section{The protocol, beta version}
\label{protocol_beta}

In this section, we give a preliminary description of  our
authentication protocol. Here Alice is the  prover and Bob the
verifier. We call this a ``beta version" of the protocol because
what we describe here represents a single session; repeating this
particular protocol several times can compromise the long-term
private key of the prover. This is why extra care has to be taken to
protect the long-term private key; this is done in the complete
protocol described in the following section, while here, in an
attempt to be helpful to the reader, we describe the ``skeleton" of
our scheme where all principal (i.e., non-technical) ideas are
introduced.

The platform ring $G$ that we suggest is the ring of $n \times n$
matrices over $N$-truncated $k$-variable polynomials over a ring
$R$. The reader is referred to our Section \ref{parameters} for the
definition of $N$-truncated polynomials as well as for suggested
values of parameters $n$, $N$, $k$, and the  ring   $R$.

\medskip

\noindent {\bf  Protocol, beta version}
\medskip

\begin{enumerate}

\item[(i)] Alice's public key is a pair of matrices $(A, X^{-1}AX)$, where the matrix $X \in G$ is Alice's long-term
private key. The matrix $A \in G$ does not have to be invertible.

\item[(ii)]  At the  challenge step, Bob  chooses a random  matrix $B$ from the ring
$G$ and sends it to Alice. (See the full version of the protocol in
the next section for how to prevent Bob from choosing $B$
maliciously.)

\item[(iii)]  Alice responds with the  matrix $X^{-1}BX$.


\item[(iv)]  Bob selects  a random  word  $w(x,y)$ (without negative exponents on $x$ or  $y$), evaluates the matrices
$M_1=w(A, B)$  and  $M_2=w(X^{-1}AX, X^{-1}BX)$, then computes their
traces. If  $tr(M_1)=tr(M_2)$, he accepts authentication. If not,
then rejects.

\end{enumerate}

The point of the final test is that $M_2=w(X^{-1}AX, X^{-1}BX)$
should be equal to   $X^{-1}M_1X= X^{-1}w(A, B)X$.  Therefore, since
the matrices  $M_1$    and  $M_2$ are conjugate, they should, in
particular, have the same trace. Note that the trace in this context
works much better (from the security point of view) than, say, the
determinant, because the determinant is a multiplicative function,
so the adversary could use any matrix with the same determinant as
$B$ in place of $X^{-1}BX$, and still pass the determinant test.
With the trace, the situation is quite different, and there is no
visible way for the adversary to pass the trace test for a random
word  $w(x,y)$  unless he/she actually uses the matrix $X^{-1}BX$.

\section{The protocol, full version}
\label{protocol_full}

Compared to the beta version described in the previous section, the
full protocol given in  this section has an extra feature of
protecting the long-term private key  $X$ from overexposure. This is
needed because upon accumulating sufficiently many matrices of the
form $X^{-1}B_iX$ with different $B_i$ but the same $X$, the
adversary may recover $X$ more easily. To avoid this, we make Alice
(the prover) apply a non-invertible endomorphism (i.e., a
homomorphism into itself) of the ambient ring $G$ to all
participating matrices. This endomorphism is selected by Bob in the
beginning of each new  session. We also note yet another extra
feature of the protocol below, namely, a (mild) ``commitment to
challenge" by the verifier (step 2(i)) preceding the actual
challenge. Recall that this is done to prevent a malicious verifier
from presenting the prover with a carefully selected challenge that
may result in leaking information about the prover's private key at
the response step.

\medskip

\noindent {\bf  Protocol, full  version}
\medskip

\begin{enumerate}

\item Alice's public key is a pair of matrices $(A, X^{-1}AX)$, where the matrix $X \in G$ is Alice's long-term  private key. The matrix
$A \in G$ does not have to be invertible.

\item  At the ``commitment to challenge" step, Bob  chooses: (i) a random  matrix $B$ from the ring $G$; (ii) a random non-invertible
endomorphism $\varphi$ of the ring $G$. Bob then sends  $B$  and
$\varphi$ to Alice.

\item In order to prevent a malicious  Bob from presenting her with a carefully selected challenge,
Alice publishes  random positive integers  $p$ and $q$ and asks Bob
to send her random non-zero constants $c_i, i=1,2,3,$ and create his
challenge in the form $B'=c_1A +c_2B +c_3A^pB^q$.

\item Upon receiving $B'$,   Alice responds with  the matrix $\varphi(X^{-1}B'X)$.

\item  Bob selects  a random  word  $w(x,y)$ (without negative exponents on $x$ or  $y$), evaluates the matrices
$M_1=w(\varphi(A), \varphi(B'))$  and  $M_2=w(\varphi(X^{-1}AX), \varphi(X^{-1}B'X))$, then
computes their traces. If  $tr(M_1)=tr(M_2)$, he accepts authentication. If not, then rejects.

\end{enumerate}

\section{Parameters and key generation}
\label{parameters}

Our suggested platform ring $G$ is the ring of all $n\times n$
matrices over  truncated $k$-variable polynomials over the  ring
${\mathbf Z}_{11}$. Truncated (more precisely, $N$-truncated)
$k$-variable  polynomials over ${\mathbf Z}_{11}$ are elements of
the factor algebra of the algebra ${\mathbf Z}_{11}[x_1, \ldots,
x_k]$ of  $k$-variable  polynomials over ${\mathbf Z}_{11}$ by the
ideal generated by all monomials of degree $N$. In other words,
$N$-truncated $k$-variable polynomials are  expressions of the form
${\displaystyle \sum_{0 \le s \le N-1} a_{j_1...j_s} \cdot x_{j_1}
\cdots x_{j_s}}$, where $a_{{j_1...j_s}}$ are elements of ${\mathbf
Z}_{11}$, and $x_{j_s}$ are variables.

To make computation  efficient for legitimate parties, we suggest to
use {\it sparse}  polynomials as entries in participating matrices.
This means that there is an additional parameter  $d$ specifying the
maximum number of non-zero coefficients in polynomials randomly
generated by Alice or Bob. Note that the number of different
monomials of degree $N$ in $k$ variables is $M(N,k) = {N+k \choose
k}$. This number grows exponentially in $k$ (assuming that  $N$ is
greater than  $k$). The number of different collections of $d$
monomials (with non-zero coefficients) of degree $<N$ is more than
${M(N,k) \choose d}$, which grows exponentially in both $d$ and
$k$. Concrete suggested  values for parameters are given below;
right now we just say that, if we denote the {\it  security
parameter} by $t$, we  suggest that the number $M(N,k) = {N+k
\choose k}$ is at least $t$. At the same time, neither $N$  nor  $k$
should exceed $t$. As for the parameter $d$, we require that
$d^{\frac{m}{n}} \cdot k \cdot \log N \cdot n^2<t$, where $m$ is yet
another parameter, defined in the following subsection
\ref{matrix_gen}.

Since the questions of  generating  random invertible matrices or
random polynomial  endomorphism have not been addressed in the
literature on cryptography before (to the best of our knowledge), we
address these questions below.

\subsection{Generating matrices}
\label{matrix_gen}

Our notation here follows that of Section \ref{protocol_full}.

Since the matrices $A$ and $B$ do not have to be invertible, they
are easy to generate.  We require that each  entry is a
$\sqrt{d}$-sparse $N$-truncated $k$-variable  polynomial over
${\mathbf Z}_{11}$, which is generated the obvious way. Namely, one
first chooses $\sqrt{d}$ random monomials of degree at most $N-1$,
then randomly chooses non-zero coefficients from ${\mathbf Z}_{11}$
for these monomials.

An invertible matrix $X$ can be generated as a random  product of
$m$ {\it elementary} matrices. A square matrix is called  elementary
if it differs from  the identity matrix by exactly one non-zero
element outside the diagonal. This single non-zero element is
generated as described in the previous paragraph. Denote by
$E_{ij}(u)$ the elementary matrix that has  $u \ne 0$ in the $(i,
j)$th place, $i \ne j$.

We note that multiplying $m$ elementary matrices  may result in the
number of non-zero  coefficients in some of the entries growing
exponentially in $m$. More precisely,  when we multiply $E_{ij}(u)$
by $E_{jk}(v)$, the result is $E_{ik}(uv)$, and the polynomial $uv$
is no longer $d$-sparse, but $d^2$-sparse. However, this phenomenon
is limited to products of elementary matrices of the form $E_{ij}(u)
\cdot E_{jk}(v)$, and the expected maximum length of such
``matching" chains in a product of $m$  elementary $n\times n$
matrices is $\frac{m}{n}$. We therefore require that
$d^{\frac{m}{n}} \cdot k \cdot \log N \cdot n^2 <t$, where $t$ is
the security parameter.

\subsection{Generating an endomorphism}
\label{endomorphism_gen}

At step 2 of the full protocol in Section \ref{protocol_full}, Bob
has to generate a random non-invertible   endomorphism $\varphi$ of
the ring $G$ of matrices over $N$-truncated $k$-variable polynomials
over ${\mathbf Z}_{11}$.

Such an endomorphism is going to be naturally induced by   an
endomorphism of the ring of $N$-truncated  $k$-variable polynomials
over ${\mathbf Z}_{11}$. The latter endomorphism can be constructed
as follows: $\varphi : x_j \to f_{j}$, where
$f_{j}=f_{j}(x_1,\ldots, x_k)$ are random sparse $N$-truncated
$k$-variable polynomials over ${\mathbf Z}_{11}$  {\it with zero
constant term}, which actually depend on $(k-k_0)$ variables only,
i.e., $k_0$ variables are missing, where the parameter $k_0$ is
specified in the following subsection.  The zero constant term
condition is needed for $\varphi$  to actually be an endomorphism,
i.e., to keep invariant the ideal generated by all monomials of
degree $N$. For efficiency reasons, it makes sense to have the
polynomials $f_{j}$ $\sqrt{d}$-sparse.

%
%
%

\subsection{Suggested parameters}
\label{parameters2}

Suggested  values for parameters of our scheme are:

\begin{enumerate}

\item The suggested  value of $n$ (the size of participating matrices) is $n=3$.

\item   Presently,  $N=1000$, $d=25$, and  $k=10$ should be quite enough to meet the security conditions specified above.
In particular, with these values of parameters, the
number $M(N,k)$ of   different monomials is greater than  $10^{20}$.

\item  The matrix $X$ (Alice's long-term  private key)  is generated by Alice as a product of
$m$ random elementary matrices, where the value for $m$ is randomly selected from the interval  $n^3 \le m \le 2n^3$.

\item  Parameter $k_0$  used in constructing a non-invertible  endomorphism (subsection \ref{endomorphism_gen} above) can be
specified as follows: $k_0$ is randomly selected from the interval  $\frac{k}{3} \le k_0 \le \frac{2k}{3}$.

\item  Values of random positive integers  $p$ and $q$ in step 3 of the protocol in Section \ref{protocol_full} can be bounded by
5.  Non-zero constants $c_i$ in the same step 3 are selected
uniformly randomly from the set of all non-zero elements of
${\mathbf Z}_{11}$.

\item The suggested   length of the word  $w(x,y)$ in step 5 of the protocol in Section
\ref{protocol_full} is 10.

\end{enumerate}

\subsection{Key size and key space}
\label{size}

To conclude this section, we point out that the size of a random
matrix in our scenario  (e.g. Bob's commitment $B$) is $\sqrt{d}
\cdot k \cdot \log N \cdot n^2$. The size of an {\it invertible}
matrix $X$ is, roughly, $(d \cdot k \cdot \log N + \log n) \cdot m$.

The size of the key space for the long-term private key (i.e., the matrix $X$) is, roughly,
$exp((d \cdot k \cdot \log N + \log n) \cdot m)$.

\section{Cryptanalysis}
\label{cryptanalysis}

We start by discussing how the adversary, Eve, can attack Alice's
long-term private key (the matrix $X$)  directly, based just on the
public key  $P=X^{-1}AX$. The relevant problem is known as the {\it
conjugacy search problem}. Note that the equation $P=X^{-1}AX$
implies $XP=AX$, which translates into a system of $n^2$ linear
equations for the entries of $X$, where $n$ is the size of
participating matrices. Thus, a natural way for Eve to attempt to
find $X$ would be to solve this system. However, there are some
major obstacles along this way:

\begin{enumerate}

\item[(a)]  The matrix equation $XP=AX$ is  {\it not equivalent} to $P=X^{-1}AX$. The former equation has many solutions; for example,
if $X$ is a solution, then any matrix of the form $X'=f(A)\cdot X
\cdot g(P)$ is a solution, too, where  $f(A)$  and  $g(P)$ are
arbitrary polynomials in the matrices $A$ and  $P$, respectively.
However, only {\it invertible} matrices $X'$ will be solutions of
the equation $P=X^{-1}AX$. If  participating matrices come from a
ring where ``generic" matrices are non-invertible (which is the case
for our suggested platform ring), then  Eve would have to add to the
matrix equation $XP=AX$ another equation $XY=I$, where $X, Y$ are
unknown matrices, and $I$ is the identity matrix. This translates
into a system of $n^2$ {\it quadratic} equations, not linear ones.

\item[(b)]  As explained in the previous paragraph,  Eve is facing a system of $n^2$ linear equations and $n^2$ quadratic equations,
with $2n^2$ unknowns, over a ring $R$, which  in our scheme is the
ring of $N$-truncated $k$-variable polynomials over ${\mathbf
Z}_{11}$. She can further  translate this into a system of linear
equations over  ${\mathbf Z}_{11}$ if she collects coefficients at
similar monomials, but this system is going to be huge: as explained
in our Section \ref{parameters}, it is going to have more than
$10^{20}$  equations (by the number of monomials). Note that,
although entries of all participating matrices are sparse
polynomials, Eve does not know which monomials in the private matrix
$X$ occur with non-zero coefficients, which means she has to either
engage {\it  all}   monomials in her equations or try all possible
supports (i.e., collections of monomials with non-zero coefficients)
of the entries of elementary matrices in a decomposition of $X$ (see
subsection \ref{matrix_gen}).

\item[(c)]  Eve may hope to get more information about the matrix $X$ if she eavesdrops on several authentication sessions between
legitimate parties. More specifically, she can accumulate several
pairs of matrices of the form $(\varphi_i(B_i),
\varphi_i(X^{-1}B_iX))$. Note however that even if a pair like that
yields some information, this is going to be information about the
matrix $\varphi_i(X)$ rather  than about $X$ itself. To recover $X$
from  $\varphi_i(X)$  is impossible because $\varphi_i$ has a large
kernel by design.

\end{enumerate}

\bigskip

\noindent {\it Acknowledgement.} Both authors are grateful to  Max
Planck Institut f\"ur Mathematik, Bonn for its hospitality during
the work on this paper.

\end{document}